\documentclass[conference]{IEEEtran}
\IEEEoverridecommandlockouts

\usepackage{cite}
\usepackage{amsmath,amssymb,amsfonts}
\usepackage{algorithmic}
\usepackage{graphicx}
\usepackage{textcomp}
\usepackage{xcolor}
\usepackage{multirow}
\usepackage{graphicx}
\usepackage{array}
\usepackage{balance}
\usepackage{hyperref}

\def\BibTeX{{\rm B\kern-.05em{\sc i\kern-.025em b}\kern-.08em
    T\kern-.1667em\lower.7ex\hbox{E}\kern-.125emX}}

\begin{document}
\bstctlcite{IEEEexample:BSTcontrol}

\title{Accelerating PoT Quantization on Edge Devices

}

\author{Rappy Saha, Jude Haris, Jos\'e Cano \\
\emph{School of Computing Science, University of Glasgow, Scotland, UK}
}

\maketitle


\begin{abstract}

Non-uniform quantization, such as power-of-two (PoT) quantization, matches data distributions better than uniform quantization, which reduces the quantization error of Deep Neural Networks (DNNs). 
PoT quantization also allows bit-shift operations to replace multiplications, but there are limited studies on the efficiency of shift-based accelerators for PoT quantization. 
Furthermore, existing pipelines for accelerating PoT-quantized DNNs on edge devices are not open-source.

In this paper, we first design shift-based processing elements (shift-PE) for different PoT quantization methods and evaluate their efficiency using synthetic benchmarks.
Then we design a shift-based accelerator using our most efficient shift-PE and propose PoTAcc, an open-source pipeline for end-to-end acceleration of PoT-quantized DNNs on resource-constrained edge devices. 
Using PoTAcc, we evaluate the performance of our shift-based accelerator across three DNNs.
On average, it achieves a $1.23\times$ speedup and $1.24\times$ energy reduction compared to a multiplier-based accelerator, and a $2.46\times$ speedup and $1.83\times$ energy reduction compared to CPU-only execution. Our code is available at \texttt{https://github.com/gicLAB/PoTAcc}

\end{abstract}



\section{Introduction}
\label{sec:introduction}

Deploying Deep Neural Networks (DNNs) on resource-constrained edge devices requires a careful balance between efficiency and accuracy~\cite{gibson_dlas_2024}. 
Ideally, the goal is to run DNN models on edge devices with floating-point accuracy while meeting performance and energy budgets.
One key technique for achieving this is quantization. 
While it is well known that 8-bit uniform quantization can achieve accuracy comparable to floating-point, previous research~\cite{li_additive_2020,chang_mix_2021,przewlocka-rus_power--two_2022} has shown that 4-bit power-of-two (PoT) quantization, a type of non-uniform quantization, can better match data distribution compared to uniform quantization and still achieve good accuracy using less bits.
Additionally, PoT quantization allows replacing multiplications by cheaper bit-shift operations. 

In general, a single bit-shift operation is computationally cheaper than a multiplication, but different PoT quantization methods introduce varying numbers of shifts and additions to replace a single multiplication.
For example, the methods from Przewlocka-Rus et al.~\cite{przewlocka-rus_power--two_2022} and QKeras~\cite{qkeras} use a single shift operation to replace a multiplication.
Other methods like APoT~\cite{li_additive_2020} and MSQ~\cite{chang_mix_2021} replace a multiplication by two shift operations and one addition.
Such variations in PoT quantization methods introduce different complexities in hardware design, including the processing elements (PEs).
Therefore, while previous works on PoT quantization~\cite{li_additive_2020,chang_mix_2021,przewlocka-rus_power--two_2022} focused on achieving accuracy comparable to floating-point, they provide limited discussion on hardware design complexity of the PEs for the different PoT quantization methods on edge devices.
Furthermore, another important gap we identified is the lack of open-source pipelines to explore different PoT quantization methods for DNNs, accelerate them using custom hardware and evaluate them on resource-constrained edge devices.

In this paper, we first design shift-based processing elements (shift-PEs) for three PoT quantization methods and evaluate their hardware efficiency using synthetic benchmarks.
Then we use our most efficient shift-PE to design a shift-based accelerator for DNN inference and 
integrate it into PoTAcc, our proposed open-source pipeline for end-to-end acceleration and evaluation of PoT-quantized DNNs on resource-constrained edge devices.
PoTAcc includes QKeras~\cite{qkeras} to explore different PoT quantization methods and define PoT quantized DNNs, and SECDA-TFLite~\cite{haris_secda-tflite_2023} to design and evaluate custom hardware accelerators for PoT quantized DNNs. 
Using PoTAcc, we evaluate the performance of our shift-based accelerator across three well-known DNNs on the ImageNet dataset. 

The contributions of this paper include the following:

\begin{enumerate}
    \item Design of three shift-based PEs for three PoT quantization methods and analysis of their efficiency in terms of latency, energy and resource utilization.
   
    \item Design of a shift-based accelerator using our most efficient shift-PE and integration into PoTAcc, our proposed pipeline for end-to-end acceleration of PoT-quantized DNNs on resource-constrained edge devices.
    
    \item Evaluation of the performance and energy efficiency of our shift-based accelerator on three DNN models (MobileNetV2, ResNet18, InceptionV1); and comparison against a multiplier-based accelerator ($1.23\times$ speedup and $1.24\times$ energy reduction) and CPU-only execution ($2.46\times$ speedup and $1.83\times$ energy reduction).  
\end{enumerate}

\section{Background and Related Work}
\label{sec:background_related work}


\subsection{Power-of-Two Quantization}

Non-uniform quantization involves unequal distances between quantization levels (in uniform quantization all distances are equal), and one form is power-of-two (PoT) quantization~\cite{miyashita_convolutional_2016}.
The simplest PoT quantization method (see Equation~\ref{eq:log2-quantization}) uses a single PoT term, \(Q(x)\), to express a quantization level that represents the original data; more complex methods can use a combination of multiple PoT terms.
PoT terms enable the use of bit-shift operations to implement multiplications, since shifting by the log of a PoT term is equivalent to multiply by it.
We refer to \(|log2(Q(x))|\) as the shift term, and the simplest PoT quantization method only needs a single shift term to implement a multiplication operation. 
\begin{equation}
\small
Q(x) = 2^{\lfloor \log_2(x) \rceil} \quad , \text{where} \quad x \neq 0 
\label{eq:log2-quantization}
\end{equation}
Previous works proposed different PoT quantization methods for DNNs to show efficacy over uniform quantization.
Some works (\!\!~\cite{miyashita_convolutional_2016,zhou_incremental_2017,elhoushi_deepshift_2021,przewlocka-rus_power--two_2022,jiang_jumping_2023}) use a single PoT term per quantization level, while others (\!\!~\cite{gudovskiy_shiftcnn_2017,vogel_bit-shift-based_2019,li_additive_2020,liu_optimize_2020,ansari_improved_2021,chang_mix_2021,gong_elastic_2021,xia_energy-and-area-efficient_2023}) can have multiple PoT terms per quantization level. 
In general, the shift terms derived from the PoT terms are represented within $2$ to $8$ bits.
A key challenge is to understand how the bit-width and the number of PoT terms allocated per quantization level affect the design of an associated hardware accelerator, and more specifically the PE design. Previous works do not provide such analysis. 

In this paper, we address this key point. We opted for 4-bit PoT quantization for weights and 8-bit uniform quantization for activations, since it is a commonly used configuration for low-bit PoT quantization methods~\cite{przewlocka-rus_power--two_2022}; this avoids the additional overhead of applying PoT quantization to the activations during inference. 
Note that previous works investigated joint activations and weights quantization~\cite{gudovskiy_shiftcnn_2017,gong_elastic_2021,xia_energy-and-area-efficient_2023}.


\subsection{Processing Element Design for PoT Quantization}

The core computational unit of a DNN accelerator is the PE that executes the multiply-accumulate operations (we call it mult-PE).
Hence, optimizing the mult-PE design can enhance the overall efficiency of the accelerator.
Shift-based accelerators for PoT quantization replace mult-PEs with shift-PEs.
For example, JLQ~\cite{jiang_jumping_2023} proposed a shift-PE design based on their jumping log quantization.
While the design is optimized, the circuit is inefficient in handling activations based on two's complement.
Liu et al.~\cite{liu_optimize_2020} proposed a BS-MUL architecture for 16-bit activations and 12-bit weights.
However, their target bit-width differs from ours (4-bit) and their circuit exhibits inefficiencies in handling negative activations.
Vogel et al.~\cite{vogel_bit-shift-based_2019} proposed a shift-PE based on two 8-bit shift terms, hence the accumulation in the PE circuit was expensive.
While none of the aforementioned works compare the hardware design of shift-PEs for different PoT quantization methods, Przewlocka-Rus et al. (\!\!~\cite{przewlocka-rus_power--two_2022,przewlocka-rus_hwAccPoT_2023}) discussed the efficiency of shift-PEs based on single and double shift terms and presented a hardware design for single shift-term PE, but did not provide a design for double shift-term PE. In our work, we provide hardware design and efficiency analysis of single and double shift-term PEs for different PoT quantization methods in Section~\ref{sec:analysis}.


\subsection{Pipelines for PoT Quantization Acceleration}

APoT~\cite{li_additive_2020} provides open-source code for PoT quantization of DNNs using PyTorch, but it does not include methods for accelerating or evaluating them. 
Similarly, Przewlocka-Rus et al.~\cite{przewlocka-rus_power--two_2022} uses PyTorch for PoT quantization of DNNs, and then converts the models to ONNX and TFLite formats for deployment on an FPGA platform. However, the code is not open-source. 
MSQ~\cite{chang_mix_2021} trains DNNs using PyTorch~\cite{paszke2017pytorch} and maps them to an FPGA using TVM~\cite{chen2018tvm} and custom versions of the Versatile Tensor Accelerator (VTA)~\cite{moreau2019hardware}. While VTA is an open-source accelerator, the design proposed in MSQ~\cite{chang_mix_2021} is not open-source.
Our proposed PoTAcc is an open-source pipeline for end-to-end acceleration and evaluation of PoT-quantized DNNs on resource-constrained edge devices.

\section{Hardware Analysis of PoT Quantization}
\label{sec:analysis}

We select three PoT quantization methods that use 4-bit for weights: one from QKeras~\cite{qkeras}, MSQ~\cite{chang_mix_2021}, and APoT~\cite{li_additive_2020}; they cover two types of PoT quantization, which helps demonstrate the capabilities of our PoTAcc pipeline.


\subsection{PoT Quantization Methods}

APoT~\cite{li_additive_2020} and MSQ~\cite{chang_mix_2021} are 4-bit quantization methods with two PoT terms, whereas the method from QKeras~\cite{qkeras} is 4-bit quantization with a single PoT term. 
Note that each PoT term can have several possible PoT values (Table~\ref{tab:pot-schemes}).
Furthermore, 4-bit methods have a maximum of 16 quantization levels, and methods with two PoT terms can create quantization levels by combining the PoT values available in them.

\begin{table}[t]
\centering
\caption{PoT quantization methods, terms and values.}
\label{tab:pot-schemes}
\begin{tabular}{|l|l|l|}
\hline
\textbf{PoT Method}       & \textbf{1st PoT Term}         & \textbf{2nd PoT Term}       \\ \hline
8\_4\_pot\_QKeras & \(\pm2^0\) to \(\pm2^7\)                       & NA                 \\ \hline
8\_4\_pot\_MSQ    & 0, \(\pm2^{-1}\), \(\pm2^{-2}\), \(\pm2^{-3}\) & 0, \(\pm2^{-1}\)   \\ \hline
8\_4\_pot\_APoT   & 0, \(\pm2^{-1}\), \(\pm2^{-2}\), \(\pm2^{-4}\) & 0, \(\pm2^{-3}\)   \\ \hline
\end{tabular}
\end{table}


\begin{figure*}[t]
 \centering
 \includegraphics[width=0.95\textwidth]{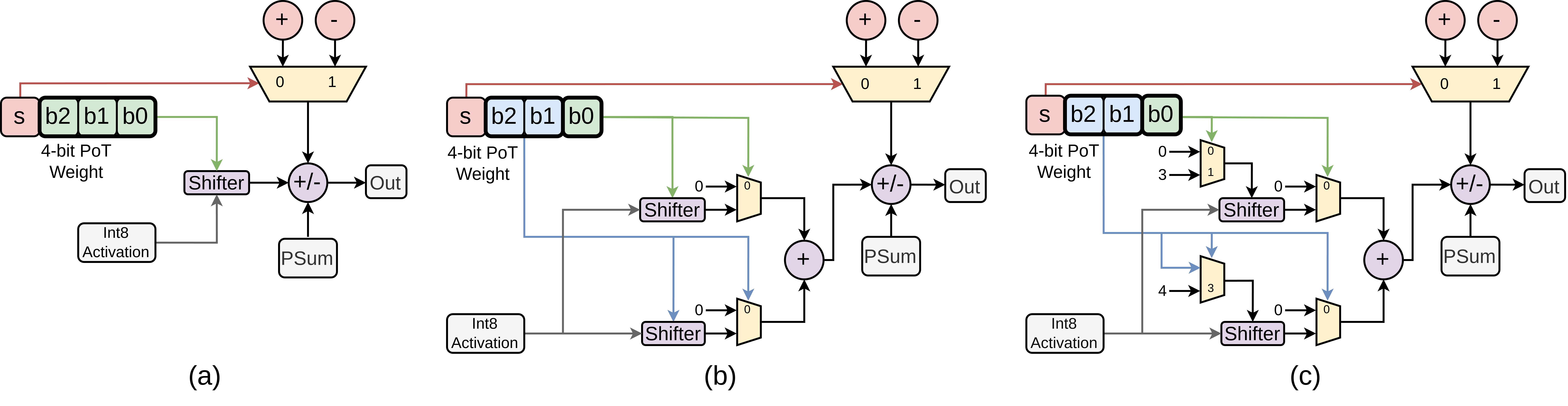}
 \caption{\label{fig:shift_pe_design} Shift-PE design: (a) 8\_4\_pot\_QKeras; (b) 8\_4\_pot\_MSQ; (c) 8\_4\_pot\_APoT.}
\end{figure*}

\subsection{Shift-PE Design}

To design a shift-PE for 4-bit PoT quantization, the left-most bit (MSB) is reserved for the sign.
Instead of storing the values of the PoT terms, we store the values of the shift terms using three bits.
For the method from QKeras~\cite{qkeras}, we assign the next three bits after MSB to the first shift term.
For APoT~\cite{li_additive_2020} and MSQ~\cite{chang_mix_2021}, the next two bits after MSB are assigned to the first shift term, whereas the right-most bit (LSB) is reserved for the second shift term.
Figure~\ref{fig:shift_pe_design} shows the shift-PE design for the three PoT quantization methods.
Note that $0$ cannot be expressed with a shift term, according to Equation~(\ref{eq:log2-quantization}). 
Besides, shifting by $0$ is equivalent to multiply by $1$.
Hence, we must treat $0$ as a special case by skipping the shift operation when the weight operand contains $0$. 
Since the method from QKeras does not have $0$ as quantization level (Table~\ref{tab:pot-schemes}), only a 3-bit shifter is required (Figure~\ref{fig:shift_pe_design}(a)).
For MSQ, since both shift terms must handle the $0$ special case, two additional multiplexers are required (Figure~\ref{fig:shift_pe_design}(b)).
In addition, when a value of a shift term cannot be represented with the designated bit width, an additional multiplexer is needed to represent it.

For example, in APoT one of the first PoT terms is \(2^{-4}\) and the shift term for it will be `$4$', which cannot be represented with the 2-bits designated for the first shift term. 
Hence, as shown in Figure~\ref{fig:shift_pe_design}(c) APoT requires an additional multiplexer to map the shift term `$3$' to `$4$'.
Note that the sign is not handled during a shift operation. 
Instead, we introduce a multiplexer to correct the sign during the addition operation.


\subsection{Hardware Efficiency of Shift-PEs}

We used Vivado high level synthesis (HLS) to design shift-PE units and analyze their resource requirement and latency.
We synthesized the designs for 250MHz, the maximum clock frequency supported by the shift-PE design in our experimental board.
Table~\ref{tab:shiftcore-hardware_cost} shows the hardware efficiency of every shift-PE design and also an 8-bit mult-PE.
We focus on the shift operation, since the accumulation operation is the same in all shift-PEs.
From Table~\ref{tab:shiftcore-hardware_cost}, we see that the method from QKeras requires both less hardware resources and clock cycles compared to the other two PoT quantization methods.

To understand the benefit of the PoT quantization method from QKeras in terms of latency and energy, we created a simple matrix-multiplication accelerator that can be reprogrammed with different multiplier/shift-based PEs.
It contains 64 PE units, and all the PE operations are done in parallel (see section~\ref{subsec:synthetic-bench} for evaluation with synthetic benchmarks).

\section{PoTAcc Pipeline}
\label{sec:pot_model_deployment}

\begin{figure} [t]
 \centering
 \includegraphics[width=0.95\columnwidth]{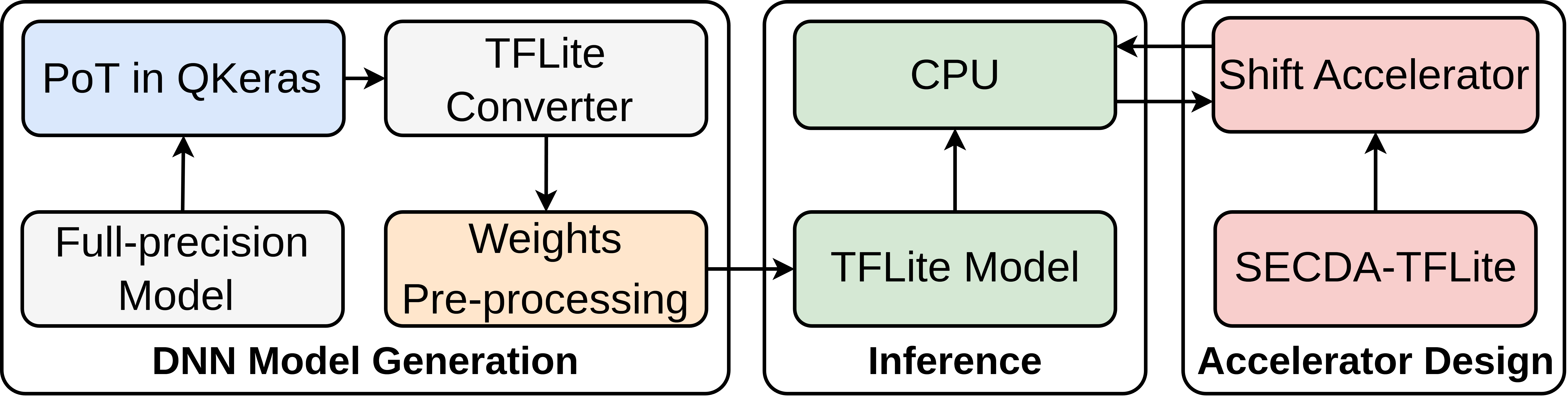}
  \caption{\label{fig:pot_deploy} PoTAcc: End-to-End Pipeline for PoT Quantization Acceleration.}
\end{figure}

We now describe PoTAcc, our open-source pipeline for end-to-end acceleration and evaluation of PoT-quantized DNNs on edge devices.
Figure~\ref{fig:pot_deploy} shows its main components: DNN Model Generation, Accelerator Design and Inference.


\subsection{DNN Model Generation}
\label{subsec:tflite_pot_inference}

Previous works~\cite{li_additive_2020,chang_mix_2021,przewlocka-rus_power--two_2022} used PyTorch to implement their PoT schemes.
Then it is possible to convert DNN models in PyTorch to TFLite~\cite{abadi2016}, but the process requires the use of custom operators for PoT quantization. 
This causes the converter to be unable to fully optimize the execution graphs of the DNN model.

We use QKeras~\cite{qkeras} to obtain PoT-quantized DNN models, since it natively supports PoT quantization.
Note that model conversion from QKeras to TFLite yields a fully optimized computational graph suited for edge inference. 
We start with a full-precision DNN model trained with TensorFlow.
We then quantize the target convolutional (\emph{conv}) layers using the QKeras PoT quantizer, and the rest of the layers are quantized to 8-bit integers.
In QKeras, it is possible to select different configurations for PoT quantization.
We select 4-bit width and maximum value of $0.5$ (i.e., PoT weights range [$-0.5$, $+0.5$]) to convert the \emph{conv} layers to quantized convolutional (\emph{qconv}) layers.
With this configuration, QKeras generates $16$ symmetric quantization levels within the [$-0.5$, $+0.5$] range.
After that, we convert the DNN model to TFLite, where weights are limited to $16$ symmetric 8-bit PoT quantization levels without zero. 
Finally, in the weights pre-processing stage, weights are converted from 8-bit PoT term format \(\pm 2^0\) to \(\pm 2^7\) to 4-bit shift term format \(\pm 0\) to \(\pm 7\).

\begin{table}[t]
\centering
\caption{Shift core hardware cost analysis based on HLS Synthesis.}
\label{tab:shiftcore-hardware_cost}
\resizebox{0.65\columnwidth}{!}{
\begin{tabular}{|l|c|c|c|}
\hline
\textbf{Quantization}      & \textbf{LUT} & \textbf{FF}  & \textbf{\#Cycles} \\ \hline
8\_4\_pot\_QKeras & 33  & 0   & 1        \\ \hline
8\_4\_pot\_MSQ    & 89  & 17  & 2        \\ \hline
8\_4\_pot\_APoT   & 118 & 19  & 3        \\ \hline
8\_8\_uni         & 41  & 0   & 2        \\ \hline
\end{tabular}
}
\end{table}

\subsection{Accelerator Design and Inference}

We use SECDA-TFLite to design our shift-based accelerator for PoT quantization based on a previous Vector MAC (VM) accelerator~\cite{haris_secda-tflite_2023}, which accelerates convolutional layers using 8-bit integers for weights and activations. The rest of the layers run on the CPU.
We replace the mult-PEs in the VM accelerator by our proposed shift-PE design, shown in Figure~\ref{fig:shift_pe_design}(a).
This requires to redesign the driver code and scheduler for efficient data packing and loading on the accelerator.
The accelerator has four GEMM units, each containing 64 MAC units.
Since the bit-width for weights is reduced by half (from 8 to 4-bit), we can accommodate twice the number of weights in the accelerator weight buffer, thereby also reducing weight 
transfer time by half.

\section{Evaluation}
\label{sec:evaluation}


\subsection{Experimental Setup}

Since we focus on edge inference, we used the PYNQ-Z2 board~\cite{pynqz2} which includes an Arm Cortex-A9 dual-core CPU and an edge FPGA.
We designed our shift accelerator using the SECDA methodology~\cite{haris_secda_2021}, and the SECDA-TFLite toolkit~\cite{haris_secda-tflite_2023} helped us integrate it in TFLite to run and evaluate PoT quantized DNNs within our PoTAcc pipeline.


\subsection{Synthetic Benchmarks}
\label{subsec:synthetic-bench}

To evaluate our shift-PE designs, we used a simple matrix-multiplication (MM) accelerator and different dimensions for the input matrices.
We send two matrices in a tiled manner with dimensions \(m \times k\) and \(k \times n\) as input, and receive the output matrix of \(m \times n\) dimensions from the MM accelerator.
We permuted the matrix sizes in the \(m\), \(n\), and \(k\) dimensions with the following values: \(m = [128, 256, 512]\); \(n = [64, 256, 1024]\); \(k = [256, 512, 1024]\) and created \(27\) synthetic benchmarks.
We removed the time required for loading and receiving data to measure computation time accurately.
As shown in Table~\ref{tab:synthetic-benchmark}, the shift-PE based on the method from QKeras~\cite{qkeras} achieves a normalized speedup of \(1.60\times\) and an energy reduction of \(1.55\times\) compared to the 8-bit integer multiplier.
It also performs well compared to other shift-PEs due to its simpler circuit design, as shown in Figure~\ref{fig:shift_pe_design}.
Even though the shift-PEs of MSQ and APoT contain multiple multiplexers, shifters, and adders due to the PoT quantization method, they are still faster than an 8-bit integer mult-PE due to the lower cost of shifters over multipliers.

\begin{table}[t]
\centering
\caption{Synthetic Benchmarks Evaluation.}
\label{tab:synthetic-benchmark}
\resizebox{0.92\columnwidth}{!}{
\begin{tabular}{|l|c|c|}
\hline
\textbf{Quantization}      & \textbf{Average Speedup}  & \textbf{Average Energy Reduction}  \\ \hline
8\_4\_pot\_QKeras & \textbf{1.60}x    & \textbf{1.55}x            \\ \hline
8\_4\_pot\_MSQ    & 1.33x             & 1.31x                     \\ \hline
8\_4\_pot\_APoT   & 1.14x             & 1.14x                     \\ \hline
8\_8\_uni         & 1.00x             & 1.00x                     \\ \hline
\end{tabular}
}
\end{table}


\subsection{End-to-end Evaluation}

We evaluated our shift-based accelerator with MobileNetV2~\cite{mbv2}, ResNet18~\cite{He2015DeepRL} and InceptionV1~\cite{szegedy2015going} DNN models 4-bit PoT-quantized with the ImageNet dataset~\cite{ILSVRC15} on the PYNQ-Z1 board.
We executed the convolutional layers on the accelerator and the remaining layers on the CPU.
The results are shown in Table~\ref{tab:end-to-end-evaluation}, where the key achievements for the shift-based accelerator are as follows: i) \(2.46\times\) speedup and \(1.83\times\) energy reduction on average against the CPU; ii) \(1.23\times\) speedup and \(1.24\times\) energy reduction on average against the VM accelerator~\cite{haris_secda_2021}; iii) the DNN model with larger convolutional layer dimensions (i.e., ResNet18) provides the highest speedup (\(1.47\times\)) and energy reduction (\(1.47\times\)) compared to the VM accelerator, since weight transfer time is reduced and more computation is done in the accelerator without transferring weights from memory.

\begin{table} [t]
\centering
\caption{\label{tab:end-to-end-evaluation} End-to-end evaluation with inference time (milliseconds, speedup) and energy consumption (joules, reduction).}
\resizebox{0.9\columnwidth}{!}{
\begin{tabular}{|c|l|c|c|c|c|}
\hline
\textbf{DNN} & \multicolumn{1}{c|}{\textbf{Hardware setup}} & \multicolumn{2}{c|}{\textbf{Total time}} & \multicolumn{2}{c|}{\textbf{Energy}} \\ \hline

\parbox[t]{16mm}{\multirow{3}{*}{\rotatebox[origin=c]{0}{MobileNetV2}}}  & CPU                         & 362           & 1.00x          & 0.45           & 1.00x          \\ \cline{2-6} 
                                                                         & VM~\cite{haris_secda_2021}  & 247           & 1.47x          & 0.40           & 1.12x          \\ \cline{2-6} 
                                                                         & Shift Acc                   & \textbf{239}  & \textbf{1.51x} & \textbf{0.38}  & \textbf{1.19x} \\ \hline
\parbox[t]{16mm}{\multirow{3}{*}{\rotatebox[origin=c]{0}{RestNet18}}}    & CPU                         & 1027          & 1.00x          & 1.16           & 1.00x          \\ \cline{2-6} 
                                                                         & VM~\cite{haris_secda_2021}  & 530           & 1.94x          & 0.81           & 1.43x          \\ \cline{2-6} 
                                                                         & Shift Acc                   & \textbf{361}  & \textbf{2.85x} & \textbf{0.55}  & \textbf{2.13x} \\ \hline
\parbox[t]{16mm}{\multirow{3}{*}{\rotatebox[origin=c]{0}{InceptionV1}}}  & CPU                         & 821           & 1.00x          & 0.93           & 1.00x          \\ \cline{2-6} 
                                                                         & VM~\cite{haris_secda_2021}  & 321           & 2.56x          & 0.50           & 1.86x          \\ \cline{2-6} 
                                                                         & Shift Acc                   & \textbf{270}  & \textbf{3.03x} & \textbf{0.43}  & \textbf{2.17x} \\ \hline

\end{tabular}
}
\end{table}

\section{Conclusion}
\label{sec:conclusion}

We designed a shift-based accelerator and proposed PoTAcc, an open-source pipeline for end-to-end acceleration and evaluation of PoT-quantized DNNs on resource-constrained edge devices.
Using PoTAcc, we evaluated our accelerator on three PoT-quantized DNNs and achieved an average speedup of \(1.23\times\) and energy reduction of \(1.24\times\) compared to a multiplier-based accelerator.
In the future, we aim to evaluate more DNN models, incorporate additional PoT quantization methods, and report end-to-end accuracy.


\section*{Acknowledgment}
\small

This work was partially supported by the UK Engineering and Physical Sciences Research Council (grants EP/T517896/1 and EP/W524359/1) and the EU Project dAIEDGE (GA Nr 101120726).


\bibliographystyle{IEEEtran}
\bibliography{bib}

\end{document}